\documentclass{aa}  
\usepackage{graphicx}
\usepackage{txfonts}
\usepackage{comment}
\usepackage{ulem} 
\usepackage{orcidlink}
\usepackage{caption}
\usepackage{dirtytalk}
\usepackage{subcaption}
\usepackage{hyperref}
\newcommand{\dgr}{$^{\circ}$}
\newcommand{\co}{$^{12}$CO~}
\newcommand{\kms}{km s$^{-1}$~}
\def\hi~{H{\sc i}}

\newcommand{\orcid}[1]{\href{https://orcid.org/#1}{\includegraphics[width=8pt]{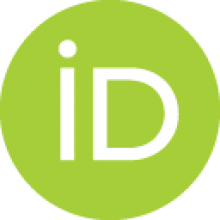}}}
\begin{document} 

   \title{The Molecular and Atomic Hydrogen Gas Content of the Bo\"otes Void galaxy CG 910}

   \titlerunning{Gas content of Void Galaxy CG 910}

   \author{Ekta Sharma \orcid{0000-0002-4541-0607}
          \inst{1,2,3}
          \and
           Prerana Biswas \href{https://orcid.org/https://orcid.org/0009-0006-7375-6580}{\includegraphics[scale=0.04]{orcid-ID.png}}
           \inst{3}
          \and 
          Mousumi Das \href{https://orcid.org/0000-0001-8996-6474}{\includegraphics[scale=0.04]{orcid-ID.png}}
          \inst{3}
          \and
          Benjamin Winkel \href{https://orcid.org/0000-0001-6999-3635}{\includegraphics[scale=0.04]{orcid-ID.png}}
          \inst{4}
          }

   \institute{National Astronomical Observatories, Chinese Academy of Sciences, A20 Datun Road, Chaoyang District, Beijing 100012, People’s Republic of China\\
              \email{sharma.ekta@bao.ac.cn, ektasharma.astro@gmail.com}
         \and
         Physical Research Laboratory, Navrangpura, Ahmedabad 380009, India
             \and
             Indian Institute of Astrophysics, Koramangala, Bangalore, 560034, India
             \and
          Max-Planck-Institut für Radioastronomie (MPIfR), Auf dem Hügel 69, 53121 Bonn, Germany
          }
            
   \date{}

 
  \abstract
   {Void galaxies are located in the most underdense environments of the Universe, where the number density of galaxies is extremely low. They are, hence, good targets for studying the secular evolution of galaxies and the slow buildup of stellar mass through star formation. Although the stellar properties of void galaxies have been studied earlier, very little is known about their cold gas content, both molecular (H$_2$) gas and atomic hydrogen (\hi~) gas.}
   {We present CO (1--0) observations of the H$_2$ gas disk in CG 910, which lies in the Bo\"otes void, which is one of the largest nearby voids and is at relatively low redshifts (z$\sim0.04-0.05$). We selected CG 910 as it is a massive disk galaxy, and early single-dish CO observations indicated that it had a large H$_{2}$ gas mass. However, the \hi~ content was not studied. Therefore, the main aim was to map the cold disk, estimate the \hi~ mass and hence the total gas mass in CG 910 and study the CO gas distribution along with the velocity field.}
   {We used the Combined Array for Research in Research in Millimeter Astronomy (CARMA) to study the CO(1--0) distribution and gas kinematics in CG 910. We also carried out atomic hydrogen observations of the galaxy using the Robert C. Byrd Green Bank Telescope (GBT). The stellar content of the galaxy and the star formation rate were derived using archival optical data.}
   {The CO(1–0) observations using CARMA reveal a molecular gas disk of mass, M(H$_{2}$) $\sim$ 12.0$\pm1.1\times10^{9}M_{\odot}$ and diameter 7 kpc. The CO velocity field shows a regularly rotating disk with a flat rotation velocity of 256 kms$^{-1}$ with no clear signatures of interaction or gas accretion. This is possibly the first CO (1--0) map of a void galaxy, and hence, important for understanding the molecular gas distribution and kinematics in void galaxies. The GBT observations reveal a \hi~ disk of mass, M(\hi~) $\sim3.1\pm0.8\times10^{9}M_{\odot}$, which is relatively small compared to the galaxy stellar mass of M$_{\star}\sim$21.5$\times$10$^{9}M_{\odot}$. The total gas mass fraction, (M(H$_2$)+M(\hi~))/$M_{\star}$ and the atomic gas mass fraction, M(\hi~)/M$_{\star}$ for CG 910 are 0.70 and 0.14, respectively.}
 {We conclude that CG 910 has a regularly rotating but massive molecular gas disk, which is more massive than the HI disk. The lower atomic gas mass fraction and star formation rate indicate a longer gas depletion timescale, confirming that, like most void galaxies, CG 910 is slowly evolving compared to normal disk galaxies.}
 
	\keywords{ISM: molecules / ISM: atoms / galaxies: star formation / Galaxy: evolution / individual objects: CG 910}
 
    \maketitle
%

\section{Introduction}
Since the early redshift surveys of galaxies \citep{1983ApJS...52...89H}, it has become clear that galaxies cluster along sheets, walls and filaments, leaving large empty regions called voids in between. Voids represent the most under-dense parts of the universe and are close to spherical, having diameters of the order of 20 to 50h$^{-1}$ Mpc for voids with less number of luminous galaxies \citep{vandeweygaert.platen.2011}. They are surrounded by walls, filaments, and clusters of galaxies, all of which form part of the cosmic web \citep{libeskind.etal.2018}. Studies have shown that voids are not completely empty but are instead populated by a distribution of late-type, low-mass galaxies that are generally gas-rich and blue in color \citep{2011AJ....141....4K,beygu.etal.2017}. They have smaller stellar masses compared to galaxies in denser environments \citep{conrado.etal.2024} and a significant fraction are blue in colour, suggesting that they are mainly late-type, star-forming galaxies \citep{2004ApJ...617...50R}. Only a small fraction show AGN activity \citep{cruzen.etal.2002,constantin.etal.2008}, which is not surprising as most void galaxies do not have prominent bulges or spheroids and do not appear to have undergone much secular evolution. In fact, most void galaxies appear to be in a relatively \lq\lq{youthful}\rq\rq state of evolution and lie in the blue part of the color-magnitude plot for galaxies \citep{2012AJ....144...16K}. 

Since void galaxies are the most underdense galaxies in the Universe, they are ideal systems for studying the slow, secular evolution of galaxies in low-density environments \citep{2011AJ....141....4K}. They also represent one of the only ways to probe the interiors of voids and are hence important for various cosmological studies related to voids, such as the redshift space distortions in voids \citep{nadathur.percival.2019} and the effect of dark energy on void evolution \citep{bos.etal.2012,vielzeuf.etal.2021}. Simulations suggest that the larger voids in the cosmic web are associated with the most massive clusters (mass $>10^{15}M_{\odot}$) \citep{shim.etal.2021} and are connected with more filaments compared to the less massive clusters (mass $<10^{14}M_{\odot}$) \citep{aragon-calvo.etal.2010}. The void galaxies in such simulations (e.g. TNG300) follow a slower evolutionary path compared to non-void galaxies, although they do undergo mass accretion via mergers at later times. They are also bluer and younger compared to galaxies in denser environments, which is similar to what is observed \citep{rodriguez.etal.2024,curtis.etal.2024}. Observations also suggest that the larger voids contain more luminous, star-forming galaxies compared to the smaller voids. A good example is the nearby large and well-studied Bo\"otes void, which contains a significant population of star-forming, emission line galaxies, some of which also host active galactic nuclei (AGN) \citep{weistrop.etal.1995,cruzen.etal.1997,pandey.etal.2021}. In contrast, the smaller voids such as the Local void contain mainly gas-rich, low luminosity galaxies such as low surface brightness (LSB) galaxies \citep{pustilnik.etal.2013} and ultra-diffuse galaxies (UDGs) \citep{roman.etal.2019}.     

In the current understanding of structure formation, voids expand at a rate faster than the Hubble flow \citep{hoffman.shaham.1982}, resulting in the clustering of galaxies near the void walls \citep{ceccarelli.eta.2012}. The void expansion thus results in smaller voids collapsing and hence merging to form larger voids \citep{sheth.vandeweygaert.2004,russell.2014}, leading to the formation of a void substructure composed of smaller voids and filaments embedded in larger voids \citep{vandeweygaert.platen.2011}. Void mergers also lead to the compression of the walls of the smaller voids and the formation of filaments within the voids. Such processes will push galaxies closer together and increase the rate of galaxy interactions and mergers, both of which lead to gas accretion and star formation. This suggests that perhaps the larger voids have the most luminous galaxies \citep{cruzen.etal.2002}.

An important question related to void galaxies is what triggers their star formation and whether it is connected to the growth of void substructures \citep{das.etal.2015}. On the scale of galaxy disks, star formation is often driven by interactions with close companions. For example, consider the galaxy pair CG 692-CG 693 in the Bo\"otes void \citep{weistrop.etal.1992} or triplets of galaxies in some nearby voids \citep{chengalur.pustilnik.2013,beygu.etal.2013}. Star formation can also be triggered by gas accretion from filaments of the cosmic web \citep{egorova.etal.2019}. The gas may flow from filaments within the voids onto the gas disks \citep{kleiner.etal.2017}. The accreted gas cools the disk and increases the gas surface density, leading to the formation of local disk instabilities and star formation. Signatures of cold gas accretion in galaxies are extra-planar gas (plumes or filaments), warped gas disks or lopsided gas distributions \citep{2008A&ARv..15..189S}. These signatures can also be detected in the \hi~~ or CO velocity fields and represent gas cooling slowly onto the disks  \citep{Boomsma.etal.2008, das_sanskriti.etal.2020}. A similar phenomenon could be happening in the low-density environments of voids; cold gas accretion from the intergalactic medium (IGM) along the filaments can drive disk star formation, resulting in the slow growth of galaxies. The neutral hydrogen (\hi~) surveys of void galaxies have detected some signatures of galaxy interactions, such as tidal tails, filaments and extended HI disks \citep{szomoru.etal.1996AJ}. 

Although void galaxies show signatures of star formation, surprisingly very little is known about their cold gas content. Some studies show that they have significant \hi~ gas masses and extended \hi~ disks \citep{2012AJ....144...16K} but there are only a few studies of the molecular gas content \cite{sage.etal.1997,beygu.etal.2013,das.etal.2015,2022A&A...658A.124D}. Studies of molecular gas disks and their gas kinematics can help one understand star formation processes in void galaxies. In this paper, we report the results of CO(1--0) observations of the molecular gas distribution and the single dish \hi~ gas observations of a void galaxy CG 910, which lies in the Bo\"otes void. These observations are slightly dated, but we think these results should be presented since so little is known about the molecular gas disks of void galaxies. In fact, this work presents the first CO gas distribution and kinematics of a disk galaxy residing in a void. The paper is organised as follows. Section 2 describes the target galaxy in detail and the reasons for selecting it. In Section 3, we describe the observations and the data reduction of CO and \hi~ emission. In Section 4, we present the results, such as gas content and morphology. We further discuss the significance of this study of the gas content in CG 910 in Section 5 and finally, we conclude the results in Section 6. 

\begin{table}
\centering
\caption{Properties of CG 910 void galaxy}\label{tab:main_galaxy}
\renewcommand{\arraystretch}{1.1}
\begin{tabular}{cccccccc} \hline \hline
Right Ascension (hh:mm:ss)    &   14:09:52.07    \\ 
Declination (dd:mm:ss)        &    +48:26:38.8    \\
Molecular Mass  (H$_{2}$)             &      3.8 $\times$  10$^{9}$ M$\odot^{\ddagger}$    \\
Redshift value        &     0.045 \\
Systematic velocity (optical) & 13135$\pm$11$^{\dagger}$ kms$^{-1}$ \\
Systematic velocity (radio) &  12600 kms$^{-1}$   \\
IRAS (100 $\mu$m) flux   & 1.56 Jy$^{\ddagger}$ \\
g band magnitude &   17.1 \\
R$_{25}$ (in arcsec, R band) & 20.97$^{\prime\prime}$ \\
Luminosity (FIR)      &    3.9 $\times$ 10$^{10}L\odot^{\ddagger}$ \\
Distance       &  188 Mpc \\
T (dust)           &     33 K$^{\ddagger}$      \\ 
Stellar mass, M$_{\star}$ [M$_{\odot}$] & 21.5 $\times$ 10$^{9}$ $(10.33)^{\star}$ \\
\hline
Properties of nearest neighbour of CG 910\\
\hline
\hline
 Right Ascension (hh:mm:ss) (J2000) & 14:10:18.5\\ 
Declination (dd:mm:ss) (J2000) & +48:25:10\\ 
 Redshift (z) & 0.043 \\
 Stellar mass, M$_{\star}$ [M$_{\odot}$] & 1.9 $\times$ 10$^{9}$ (9.84) \\
 SFR [M$_{\odot}$ yr$^{-1}$] & -0.098 \\
 Specific SFR, log sSFR [yr$^{-1}$] & -9.77 \\ 
\hline
\end{tabular}
\renewcommand{\arraystretch}{1}
$^{\dagger}$From CO spectra
$^{\ddagger}$Values taken from \cite{sage.etal.1997}.\\
$^{\star}$Value in parentheses are logarithmic scale. 
\end{table}

\begin{figure}[ht!]
\centering
\includegraphics[height=5cm, width=7cm]{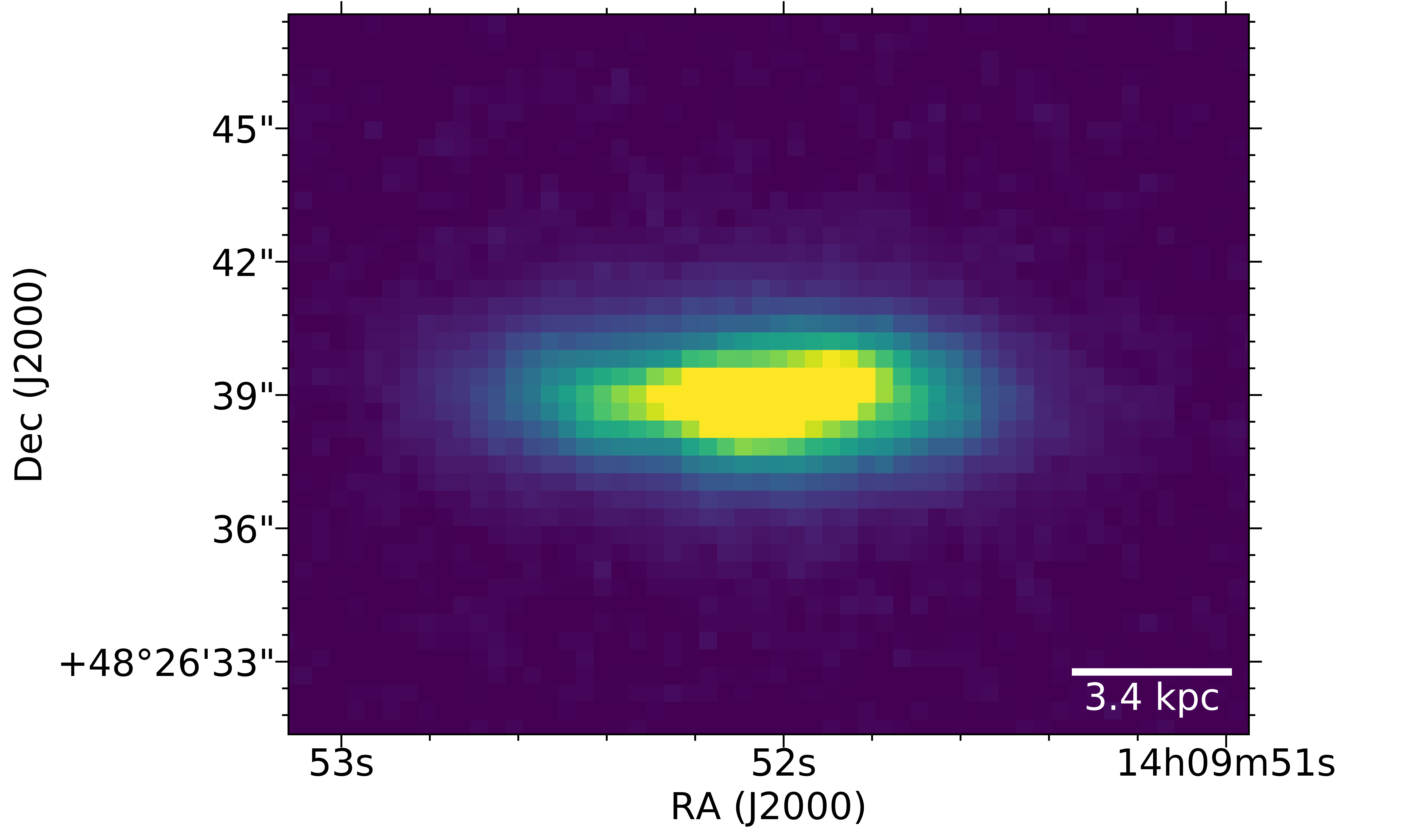}
\caption{SDSS r-band image of CG 910.}\label{fig:1_FIG}
\end{figure}
\section{The target galaxy CG 910}
We selected CG 910 as it is one of the few void galaxies that has been detected in CO emission and has a high CO flux density \citep{sage.etal.1997}. The galaxy lies within the Bo\"otes void, a large nearby void (size$\sim$60 Mpc) with a relatively large population of bright galaxies \citep{cruzen.etal.2002}. Fig. \ref{fig:1_FIG} shows the archival Sloan Digital Sky Survey (SDSS; \citealt{2000AJ....120.1579Y}) r-band emission map of CG 910. It has a relatively large optical disk of diameter $\sim$ 21$^{''}$ or 17 kpc, a prominent bulge and ongoing nuclear star formation, which is evident from its nuclear SDSS spectrum and centrally concentrated H$\alpha$ emission. The latter is evident in the SDSS-MANGA map of the galaxy. Using the SDSS H$\alpha$ flux value of $1.044\times10^{-14}$ ergs cm$^{-2}$ s$^{-1}$, distance as 188 Mpc and the star formation rate (SFR) relation from \citet{kennicutt.etal.1994}, we obtain SFR=0.33 M$_{\odot}yr^{-1}$. Hence, CG 910 has a fairly moderate SFR, but it is relatively strong compared to the SFR of other void galaxies \citep{2022A&A...658A.124D}. 

The galaxy shows radio continuum emission at 1.4 GHz in the NVSS image with a flux density of $\sim$ 4.7 mJy/beam and also in the FIRST image, where it appears as a compact core that is associated with the nucleus. There is no prior \hi~ observation of CG 910, but the CO(1–0) line has been observed using the IRAM 30 m telescope \citep{sage.etal.1997}. Table \ref{tab:main_galaxy} summarizes the properties of CG 910. The stellar mass is the best-fit value of the galaxy derived by fitting stellar evolution models to the SDSS photometry using the Portsmouth method and assuming the Kroupa Initial Mass Function (IMF). The galaxy has very few neighbours; its nearest neighbour (WISEA J141018.54+482510) is at a projected distance of 229.4 kpc and also at a redshift similar to that of CG 910 (which we determined using the NED Cone search tool). The properties of the closest neighbour are also summarised in Table \ref{tab:main_galaxy}. Thus, CG 910 appears to be a spiral galaxy with a prominent bulge and a relatively high molecular gas content, showing signatures of ongoing star formation.

\section{Observations and data reduction}
\begin{table}
\renewcommand{\arraystretch}{1.}
\setlength{\tabcolsep}{1pt}
\centering
\caption{Observation details}
\label{tab:core_pa}
\begin{tabular}{ccc}
     \hline
       &  \co & \hi~ \\
     \hline
     Observing Telescope & CARMA & Green bank \\
     Date of Observation &   June, 2014    & February, 2016  \\
     On-source time      & 2 Hours & 10 Hours \\
     Observing Frequency & 110. 26 GHz &  1360.70 MHz \\
     Channel width       & 81 kms$^{-1}$ &  20 kms$^{-1}$  \\
     Bandwidth          & 470 MHz & 1265 MHz  \\
     RMS noise$^{\dagger}$  & 4.96 mJy/beam & 0.5 mJy \\ 
     Beam size  & 3.5$^{\prime\prime}$ $\times$ 2.4$^{\prime\prime}$ $^{\dagger \dagger}$ &  9$^{\prime}$ \\   
     \hline
     $^{\dagger}${\footnotesize{Measured in line-free channel}} \\
     $^{\dagger \dagger}${\footnotesize{Synthesized beam size}}
\end{tabular}
\end{table}

\subsection{$^{12}$CO (1-0) CARMA data}
The $^{12}$CO (1-0) observations of CG 910 were made using the CARMA array of 15 antennas in the D-band configuration with only the nine 6 m BIMA dishes and six 15 m OVRO dishes. The actual bandwidth of the correlator window containing the CO line is 470 MHz with 15 channels with a velocity resolution of 81 \kms. The observations were carried out for a duration of 3.5 hours during June 2014 with an integration time of 2 hours on CG 910. The visibility calibrator is 1419+543; the bandpass calibrator and the flux calibrator were 3C84 and Mars, respectively. The data reduction package MIRIAD \citep{1995ASPC...77..433S} was used to reduce the interferometric data. A robust weighting was used with the robust parameter of 0.5.  The angular resolution of the beam is 3.1$^{''}\times2.4^{''}$ with position angle $\sim$ 90$^{\circ}$. The rms noise in the channel maps is 4.4 mJy/beam.

\subsection{\hi~ Green Bank Telescope data}
In order to study the atomic content of the galaxy, we first made \hi~ observations of CG 910 using the Effelsberg Radio telescope in June and July 2015. We applied a running median-filter-based flagging algorithm with subsequent interpolation to mitigate narrow-band Radio Frequency Interference (RFI). The flux density calibration was done following \cite{2012A&A...540A.140W}. The galaxy systemic velocity is in the radio convention and is v=12583 \kms (where v(radio)=$(\lambda-{\lambda_{0}})/\lambda$ and v(optical)=$(\lambda-{\lambda_{0}})/\lambda_{0}$). However, there is strong radio interference (from a nearby radar at 1360 MHz) in the band, which overlaps with the potential \hi~ emission from the galaxy. This emission at the high-frequency side of this RFI could be \hi~ emission from CG 910, and so further observations were needed to confirm the presence of \hi~.

In order to study the atomic content of the galaxy, we made \hi~ observations of CG 910 using the 100-m Robert C. Byrd Green Bank telescope (GBT) located in West Virginia. The observations were done in February 2016 and were carried out for a duration of 10 hours spread over two days. The back-end instrument, VEstalise GB Astronomical Spectrometer (VEGAS) in L-band in the frequency range of 1.15-1.73 GHz was used for the spectroscopic observations. The bandwidth is 1500 MHz (effective bandwidth is 1265 MHz) with a channel resolution of 1465 kHz. The position-switching mode was used. The GBTIDL software \citep{2013ascl.soft03019M} was used for the calibration and processing of \hi~ data. The noise achieved in rms after the calibration is 0.48 mJy.

\section{Results} \label{sec:highlight}
\begin{figure*}
    \centering
    \includegraphics[width=0.5\textwidth]{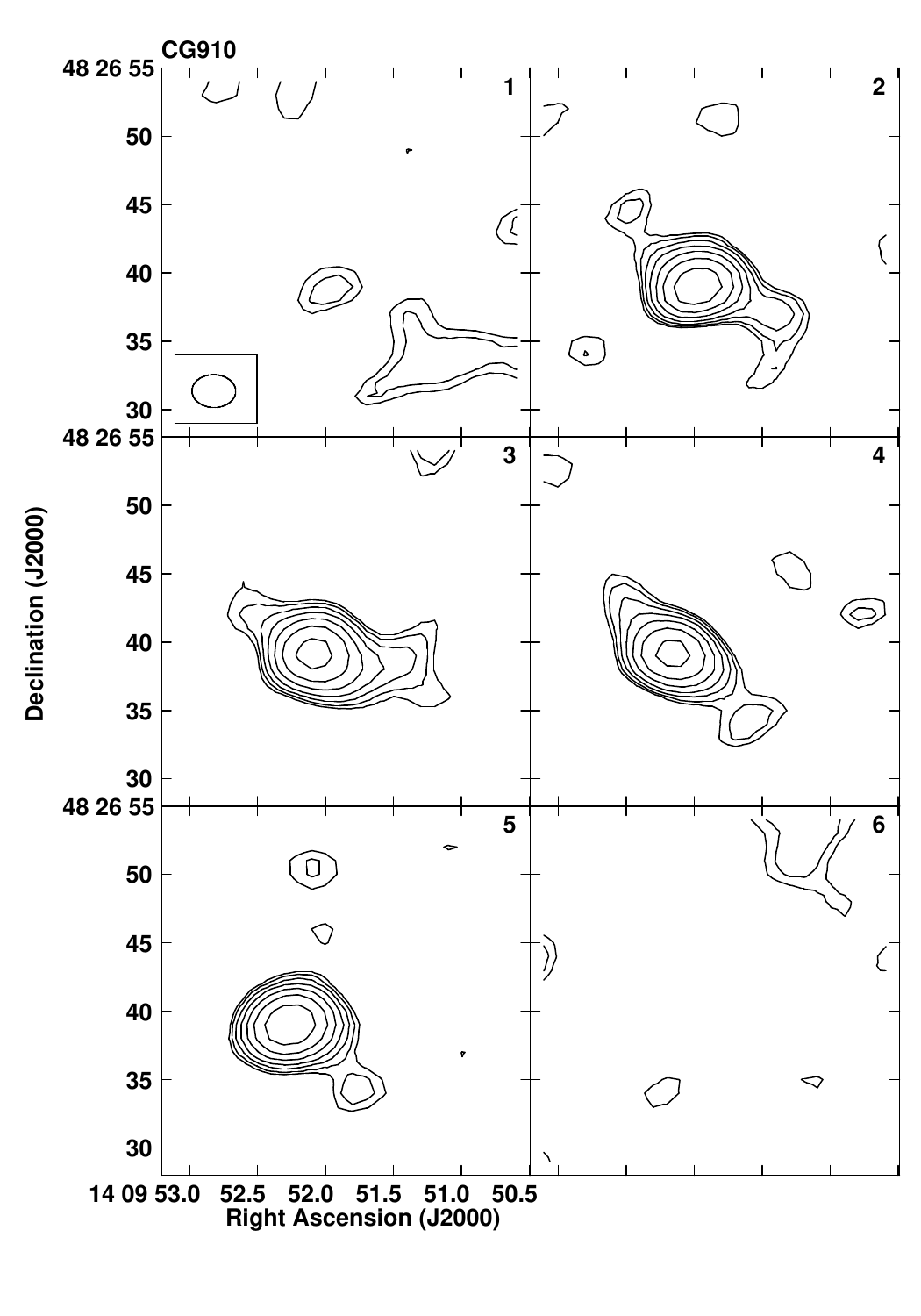}
    \caption{Channel maps for CG 910 using \co (1-0) data. The velocity in the first panel is 12746.3 \kms, with a velocity resolution of -81.2 \kms. Contour levels are set at $6.5 \times (1, 1.4, 2, 2.8, 4, 5.6, 8, 11.2, 16, 22.4, 32, 44.8, 64)$ mJy/beam, where $6.5$ mJy/beam represents the RMS noise at a line free channel of the data cube.}
    \label{fig:channel_co}
\end{figure*}

\begin{figure*}
    \centering
     \begin{subfigure}[b]{0.3\textwidth}
        \centering
        \includegraphics[width=\textwidth]{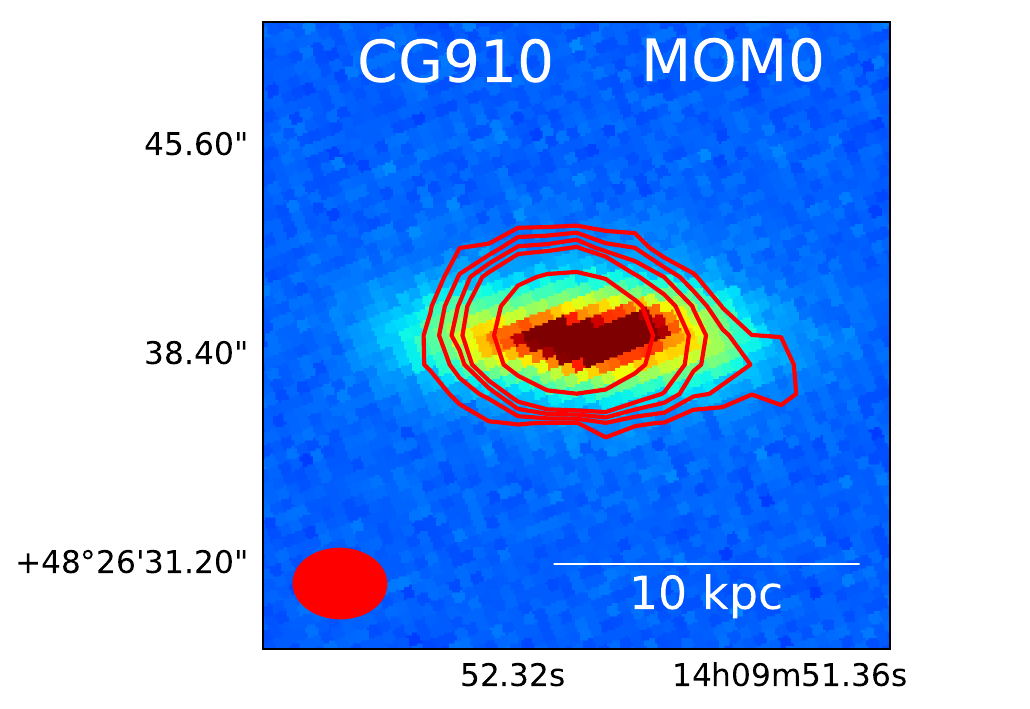}
        \caption{}
        \label{fig:co_mom0}
    \end{subfigure}
     \begin{subfigure}[b]{0.33\textwidth}
        \centering
        \includegraphics[width=\textwidth]{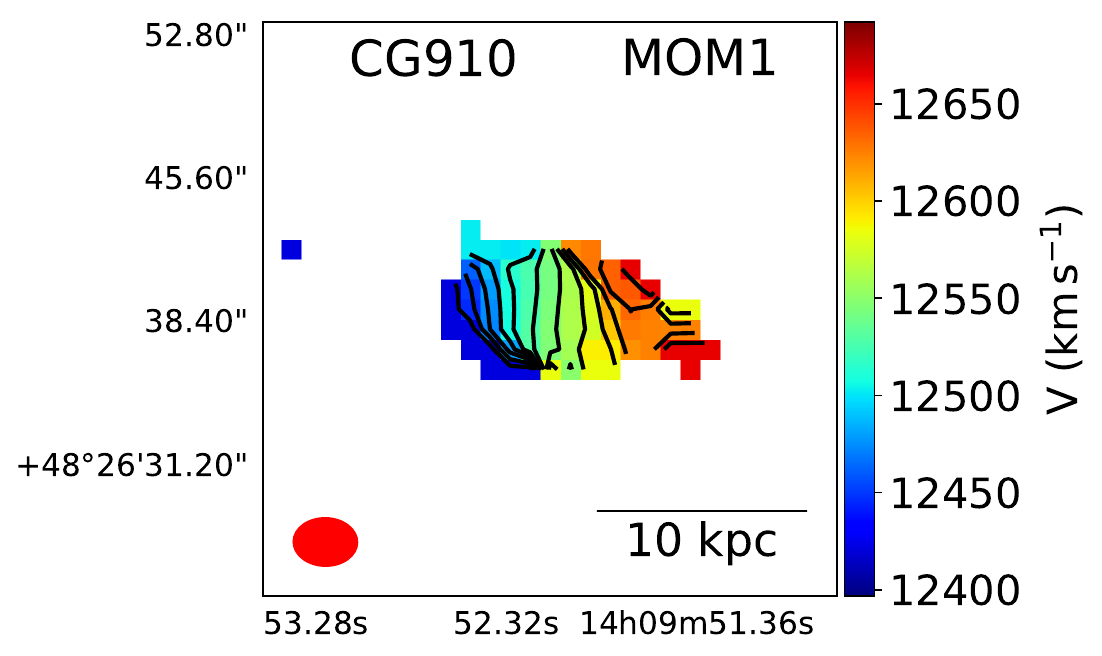}
        \caption{}
        \label{fig:co_mom1}
    \end{subfigure}
     \begin{subfigure}[b]{0.33\textwidth}
        \centering
        \includegraphics[width=\textwidth]{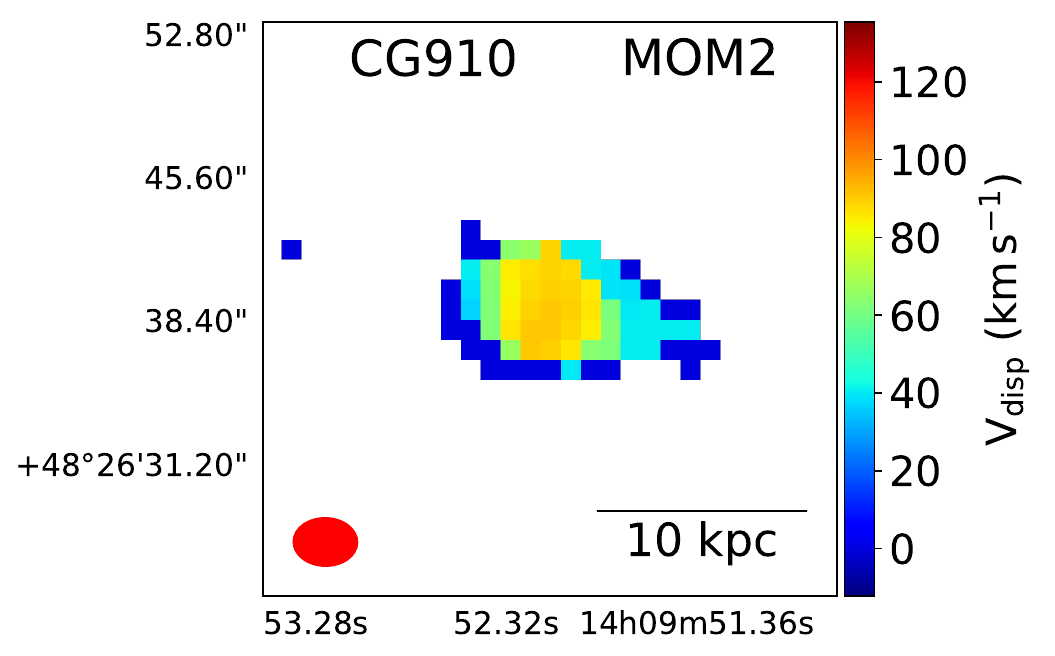}
        \caption{}
        \label{fig:co_mom2}
    \end{subfigure}
    \caption{$^{12}$CO (1-0) moment maps of CG 910. In figure \ref{fig:co_mom0}, the red contours, which represent the $^{12}$CO emission, are plotted over the SDSS r-band image. The contours are set at $(1, 2, 4, 8 ...)$ mJy/beam-km/s sigma levels. In figure \ref{fig:co_mom1}, contours are set at 20 \kms intervals. The red patch at the bottom left corner of each image represents the beam of the  $^{12}$CO map.}
    \label{fig:CO_maps}
\end{figure*}
\subsection{\co emission}
\label{subsec:channel}
The $^{12}$CO (1-0) line emission from galaxies traces the molecular gas component and is usually more centrally concentrated than the \hi~ gas distribution. To study the intensity distribution in the different channels, we made channel maps in the velocity range from 12746.3 \kms to 12339.9 \kms as shown in Fig. \ref{fig:channel_co}. The channel maps are made using the standard routine, \say{KNTR} in AIPS. Emission from the galaxy is detected in four consecutive channels (from panels 2 to 5). In each of these four channels, we observe the signature of a possible extended emission located in the southwest corner of the galaxy. The probable existence of this extended emission is based on 1-sigma detection. However, it can also be observed with 2-sigma confidence in panels 2 and 4.

Figure \ref{fig:CO_maps} represents the \co moment maps. These moment maps are made using the standard routine, \say{MOMNT} of AIPS following the procedure described in \citet{biswas2022}, i.e.,  only those channels where the line is seen in the data cube, have been used for making the moment maps. This includes four channels whose velocities range from 12746.3 \kms to 12339.9 \kms. We then measured the RMS noise from a line-free channel, $4.961$ $mJy/beam$ and used a cutoff of 1.4 times the RMS noise to make the moment maps. Further details of this procedure are described in \citet{biswas2022}. Figure \ref{fig:co_mom0} represents the integrated \co emission i.e., the moment zero map (the red contours), plotted over the SDSS r-band image. The molecular gas disk, traced by \co emission, is $\sim$ 14$^{''}$ in diameter or 13 kpc. In the moment zero map, we see a signature of the extended \co emission in the southwest direction of the galaxy detected with a 2-sigma confidence level. Figure \ref{fig:co_mom1} shows the velocity field or the moment one map of the galaxy. The gradient of the velocity field, which is a result of the galaxy's rotation, is clearly seen in this figure. The velocity of the molecular disk ranges from $\sim$ 12400 \kms to $\sim$ 12700 \kms.  Figure \ref{fig:co_mom2} shows the velocity dispersion map or the moment two map of the galaxy. The velocity dispersion map shows that the centre of the molecular disk has a high-velocity dispersion of $\sim$ 90 \kms and gradually decreases towards the end of the molecular disk to a value $\sim$ 10 \kms.

\begin{figure}[]
\centering
\includegraphics[width=0.5\textwidth]{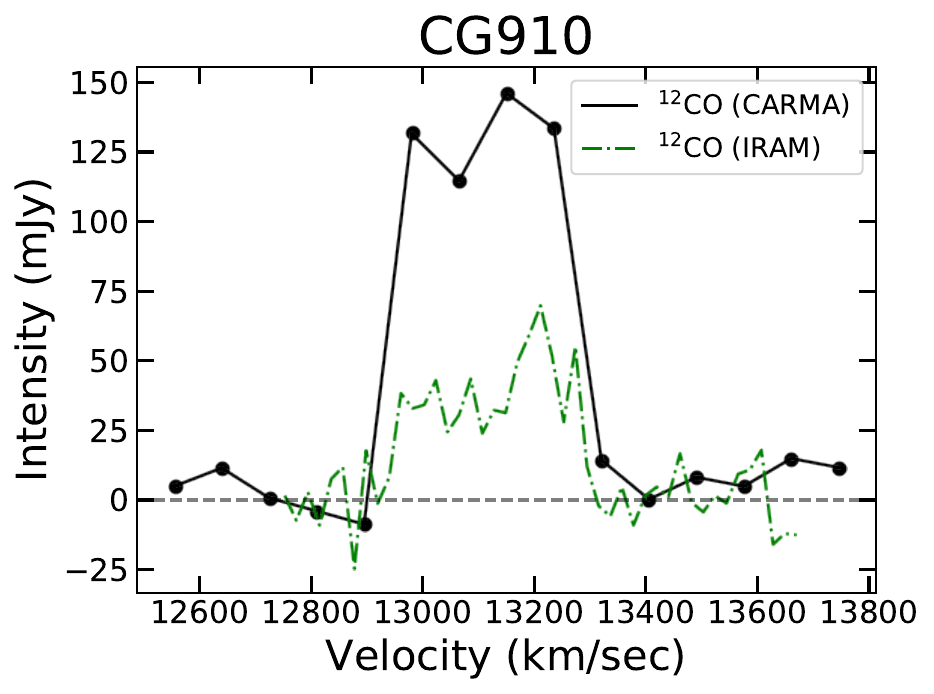}
\caption{\co spectra extracted from CARMA observation (in black) and IRAM 30 m single dish observations \citep{sage.etal.1997} (in green).}\label{fig:av_CO}
\end{figure}

\subsection{CO line profile and molecular hydrogen gas mass}
We further examined the \co line profile obtained from the CARMA data. To extract the \co spectra from the CARMA data cube, we first masked the data cube using the moment zero map of \co. In order to do that, we first draw a region around the moment zero map where the emission is seen and blank all the other regions using the standard task, \say{BLANK} of AIPS. This blanked moment zero map is further used to mask the 3-dimensional data cube so that in each channel of the data cube, only the region defined by the moment zero map has data and all the other regions are blanked. This is also been done using the \say{BLANK} task of AIPS. This masked cube is then used to extract the spectra using the task \say{ISPEC} of AIPS \citep{biswas2022}. The obtained spectrum is shown in figure \ref{fig:av_CO}.  The figure also shows a comparison of the \co spectra obtained in the IRAM 30-meter single-dish observation. The single-dish \co spectrum has been kindly provided by \citet{sage.etal.1997} through private communication. The original spectrum from \citet{sage.etal.1997} represented the brightness temperature in units of Kelvin. We converted this brightness temperature (K) to flux density (Jy) considering the IRAM beam following the procedure mentioned in \citet{Alatalo2011}. 

From Figure \ref{fig:av_CO}, we find that the \co spectra from CARMA differ from IRAM significantly, mostly in terms of amplitude. However, this large discrepancy in spectra between  CARMA and IRAM is not new.  \citet{Alatalo2013_atlas3dxviii} found that for $\sim$ 50 \% of their total sample, which includes 30 galaxies, CARMA fluxes are higher than that obtained from IRAM. \citet{Alatalo2013_atlas3dxviii} clearly mentioned that if the angular extent of the molecular gas disk, mainly the diffuse structures with a low signal-to-noise ratio such as rings, spiral arms, tidal features, etc., extends beyond a radius of 12 $^{\prime\prime}$, it can be missed by the beam of the 30-meter dish of IRAM. In the case of CG910, the radius of the molecular disk is found to be 7.1$^{\prime\prime}$.  This radius is found from the semi-major axis of an ellipse fitted to the one-sigma contour of the moment zero map from CARMA \co observation. However, we have also found extended emission associated with the galaxy through the position-velocity (PV) diagram extracted at different position angles with diameter $\sim$ 14$^{\prime\prime}$ (see figure \ref{fig:pv_co}, details about the PV diagram are mentioned in the later sections of the paper). This extended emission anticipated from the PV diagram is the most probable reason behind the discrepancy in the spectra between the CARMA and IRAM observation. It is to be noted that similar phenomena have been observed in the study of \citet{Alatalo2013_atlas3dxviii} also, for example, for NGC 1222 and NGC 4119, the emission in PV diagram extends beyond a radius of 12$^{\prime\prime}$ but the radius roughly estimated from the moment zero maps, are in between $\sim$ 7$^{\prime\prime}$ to 9$^{\prime\prime}$. Both of them have significantly higher flux in their CARMA spectra than that in IRAM spectra.

From the CO spectrum obtained from the CARMA observation, we further measure the molecular gas mass of the galaxy. To do that, we first find the line-flux of the \co line. For that, we first take the RMS noise from the line-free channel of the \co spectrum. This RMS noise is then subtracted from the whole spectrum. This has been done to remove any additional continuum emission that can still be there even after the continuum subtraction of the data cube. Then, we draw a line at the 3 times the RMS above the spectra to find out the channel range of the integration to obtain the \co line-flux. The obtained line flux is then used to find out the \co luminosity using the equation mentioned below \citep{2005ARA&A..43..677S}: 
\begin{equation}
\centering
L_{CO} [K km s^{-1} pc^{2}] =3.25\times 10^{7} S_{CO} \Delta V D_{L}^2 (\nu_{obs})^{-2} (1+z)^{-3}, 
\end{equation}
where $\Delta$V is the velocity resolution of the \co data in kms$^{-1}$, S$\Delta$V is the integrated line-flux in Jy kms$^{-1}$, $\nu$$_{obs}$ is the observed frequency in GHz which is 110.44 GHz, D$_{L}$ is the luminosity distance and $z$ is redshift of the galaxy. From the obtained \co luminosity, we calculate the molecular hydrogen gas mass of the galaxy using the following equation \citep{2013ARA&A..51..207B,2022A&A...658A.124D}: M(H$_{2}$) = $\alpha_{CO}$ L$_{CO}$ where, $\alpha_{CO}$ is the CO-to-H$_{2}$ conversion factor taken as 3.2 M$\odot$ (K km s$^{-1}$ pc$^{2}$)$^{-1}$ \citep{2013ARA&A..51..207B,2022A&A...658A.124D}. Finally, the molecular gas mass (M$_{mol}$) of the galaxy is obtained with the correction for helium contribution by taking $\alpha_{CO}$ = $4.6$. The integrated \co line-flux, derived \co mass and the molecular gas mass are mentioned in table \ref{tab:flux_mass}. The measured \co flux and the molecular hydrogen gas mass are also compared with the IRAM \co spectra from  \cite{sage.etal.1997}.  It is to be noted that the \co line-flux and the mass mentioned in the study of \citet{sage.etal.1997} are little different from what we noted in table \ref{tab:flux_mass}. This is because the distance used by \citet{sage.etal.1997} is different from the distance used by us. 

\subsection{Position-velocity diagram}
\label{subsec:pv_dia}
Since we found some indication of extended emission in the channel maps (figure \ref{fig:channel_co}) and from the moment zero map (figure \ref{fig:co_mom0}), we further investigated the position-velocity (PV) diagram of the galaxy. In order to do so, the position angle is first calculated using the following procedure. An ellipse was fitted to the one-sigma contour level of the moment zero map of the galaxy. The position angle of the galaxy is the angle measured anti-clockwise from north to the major axis of the ellipse, which is found to be 77\dgr. 

\begin{figure*}
    \centering
      \begin{tabular}{cc}
      \centering
             \includegraphics[width=0.4\textwidth]{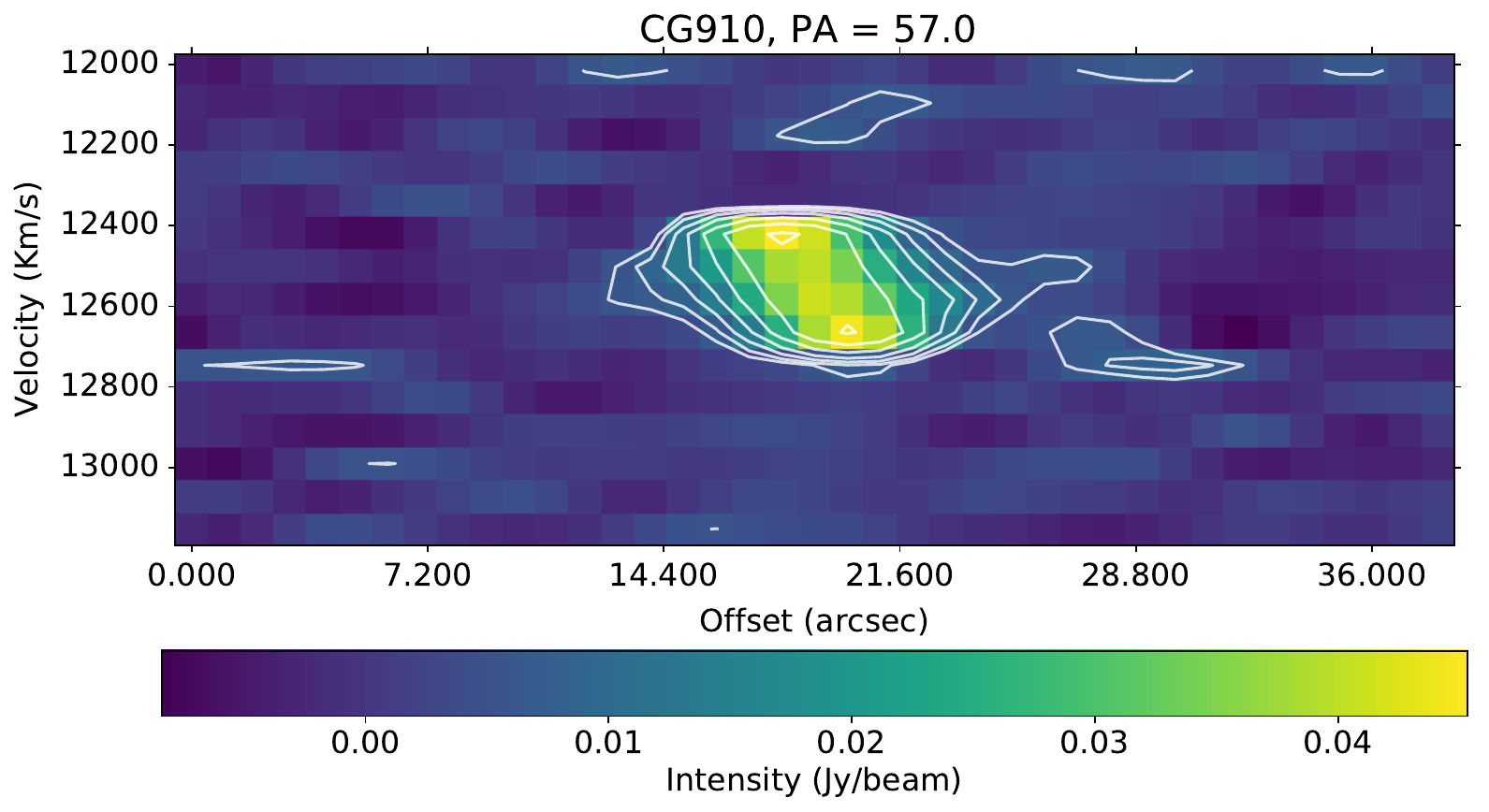} &
             \includegraphics[width=0.4\textwidth]{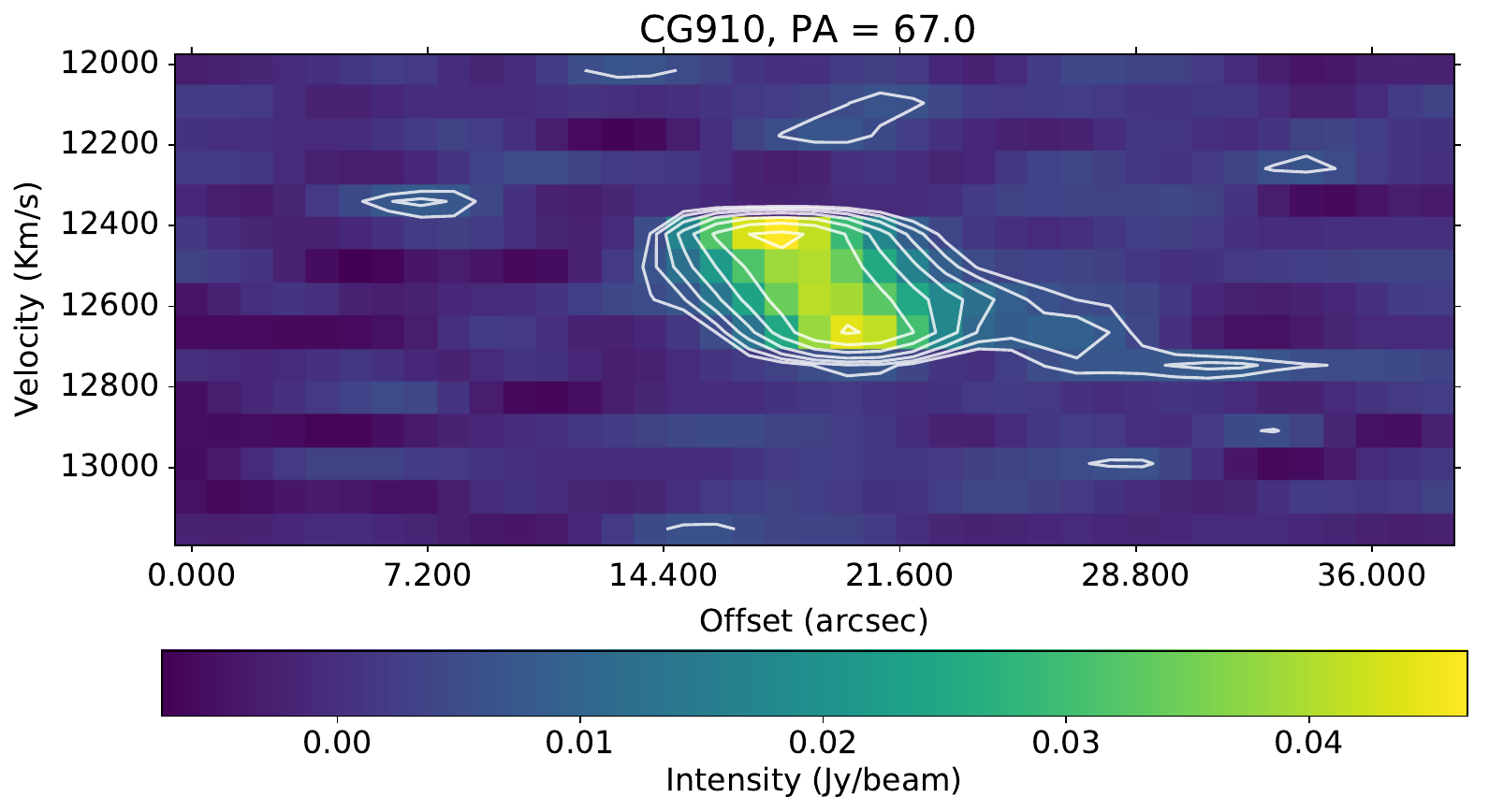} \\
             \includegraphics[width=0.4\textwidth]{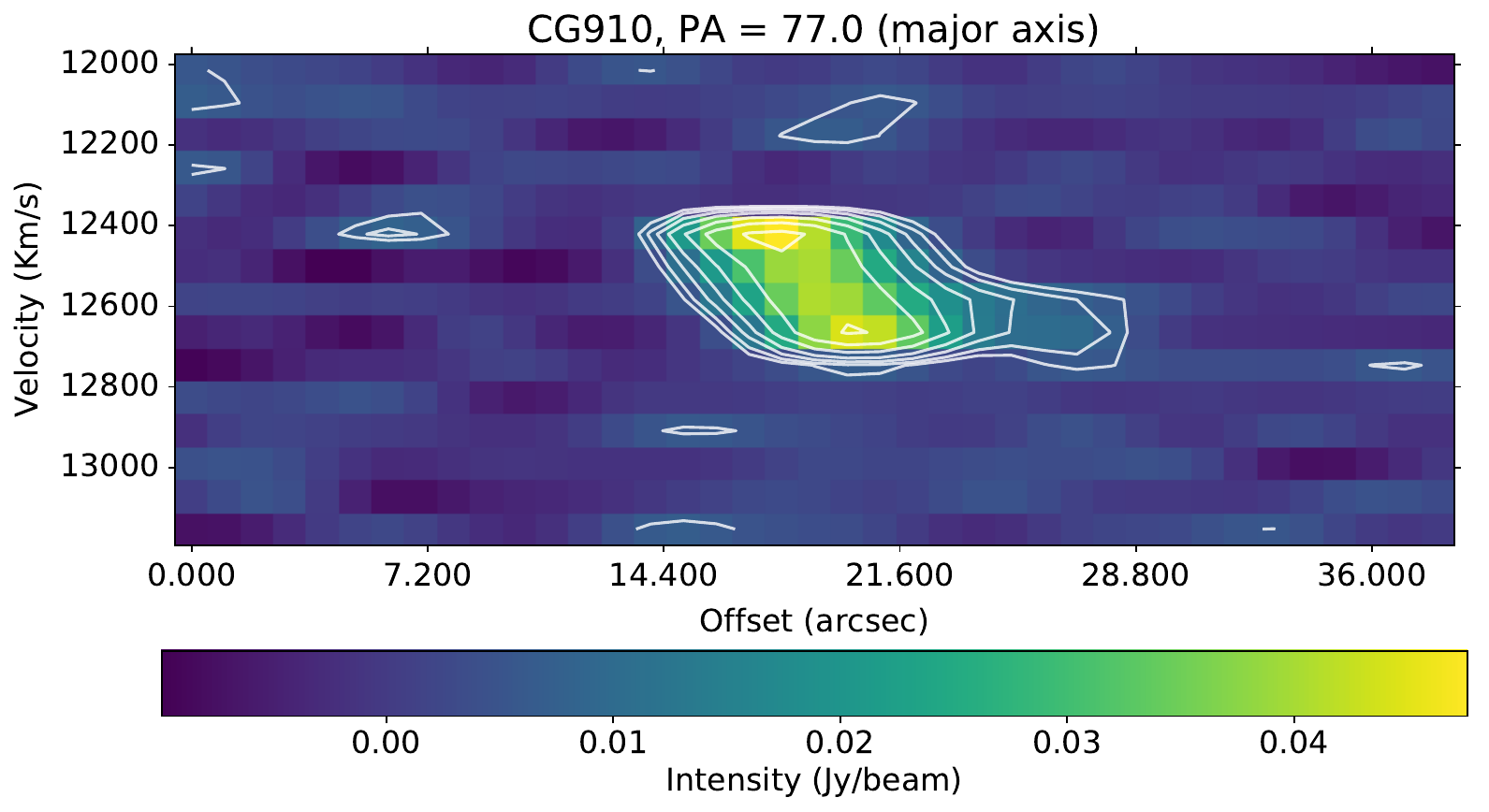} &
             \includegraphics[width=0.4\textwidth]{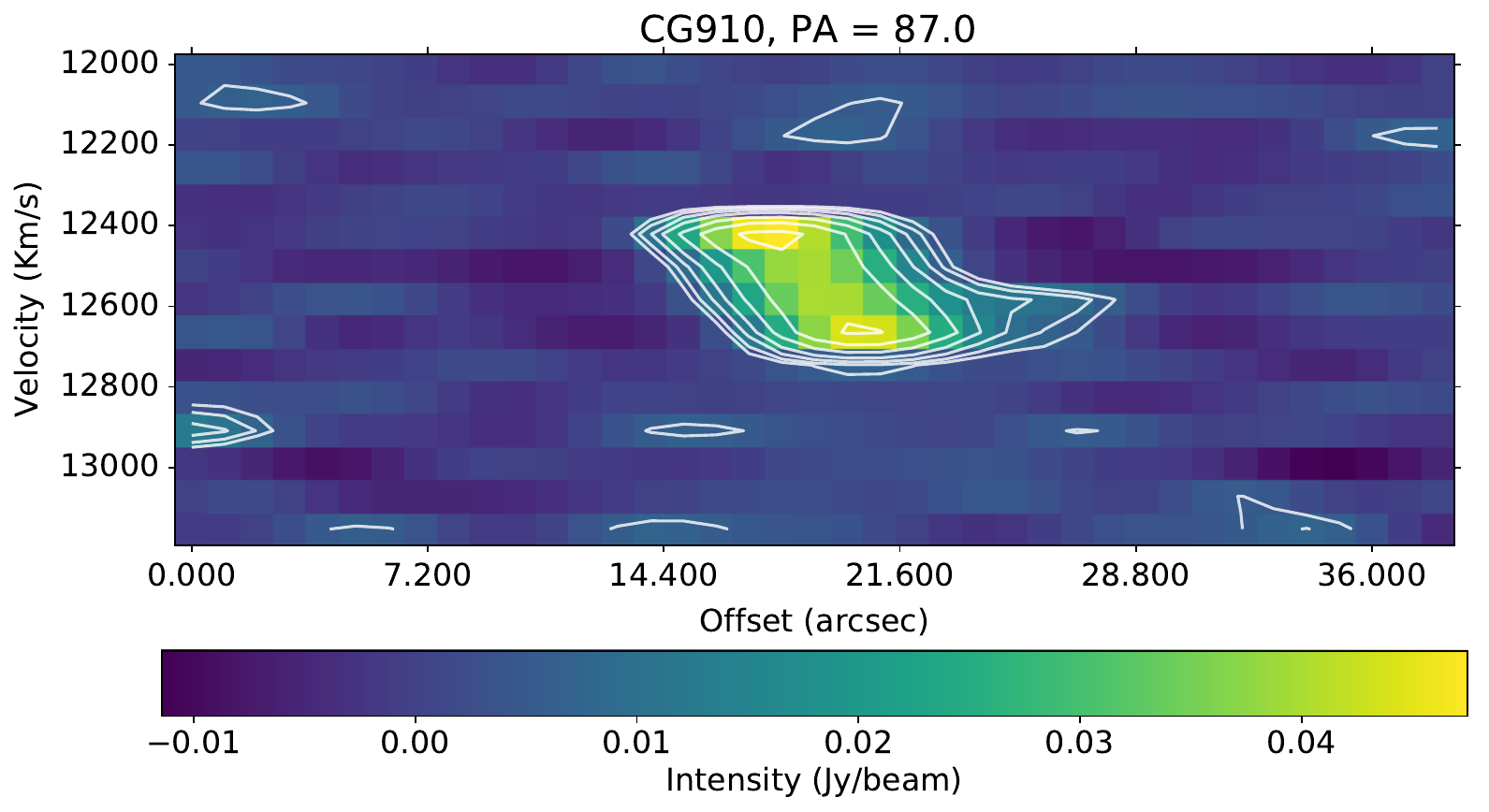} \\
             \end{tabular}
             \begin{tabular}{c}
             \centering
             \includegraphics[width=0.4\textwidth]
             {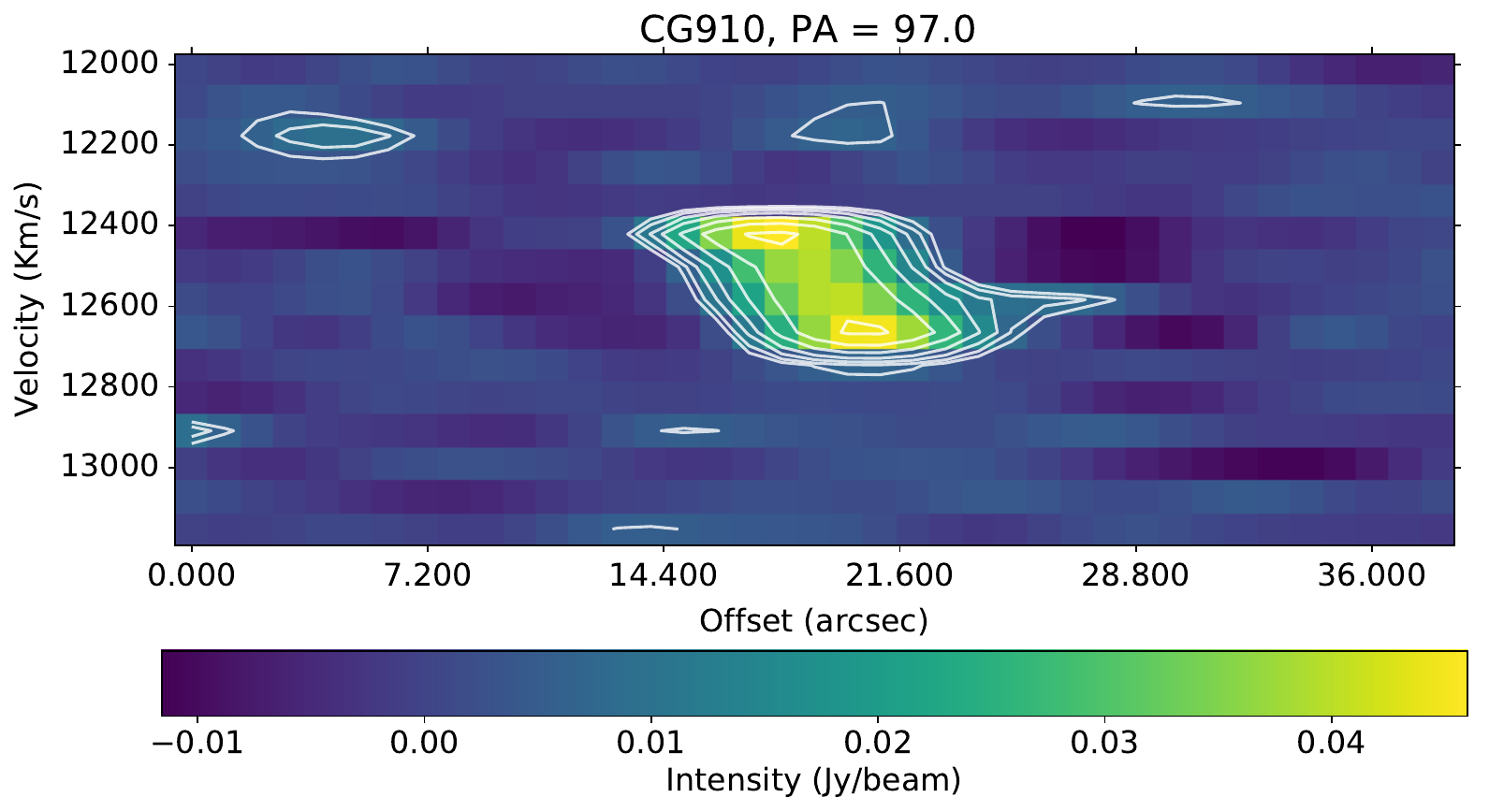} 
         \end{tabular}
         \caption{The position-velocity diagram along different slices of the galaxy. The white contours are drawn at RMS$\times(1, 1.4, 2, 2.8, 4, 5.6, 8, ...)$ levels, where the RMS is measured from a line-free channel of the \co data cube.}
    \label{fig:pv_co}    
\end{figure*}

We first extracted the PV plot along this major axis and found a signature of the extended emission detected up to a 2 sigma confidence level. We further extracted the PV plot along PA $\pm$ 10 \dgr and PA $\pm$ 20 \dgr angles. Figure \ref{fig:pv_co} represents these PV plots at each of the angles. In each of the PV plots, we find a similar signature of extended emission detected up to a 2 sigma confidence level. Hence, there may be extended emission within the $\pm$ \dgr~ of the galaxy position angle, which is 77\dgr measured counterclockwise from the north.

Most of the emission follows the typical rotation curve profile for a galaxy, with a steeply rising inner part and a flatter outer part. From the resulting PV plot shown in Figure \ref{fig:pv_co}, we find that the observed disk rotation velocity is $v_{obs}\sim$200 kms$^{-1}$. Thus, most of the molecular gas is concentrated in a 7 kpc gas disk rotating with a flat disk rotation velocity of approximately $v_{obs}$/$\sin(70^{\circ})$=256 kms$^{-1}$. 

\begin{figure}[hbt!]
\centering
\includegraphics[width=0.5\textwidth]{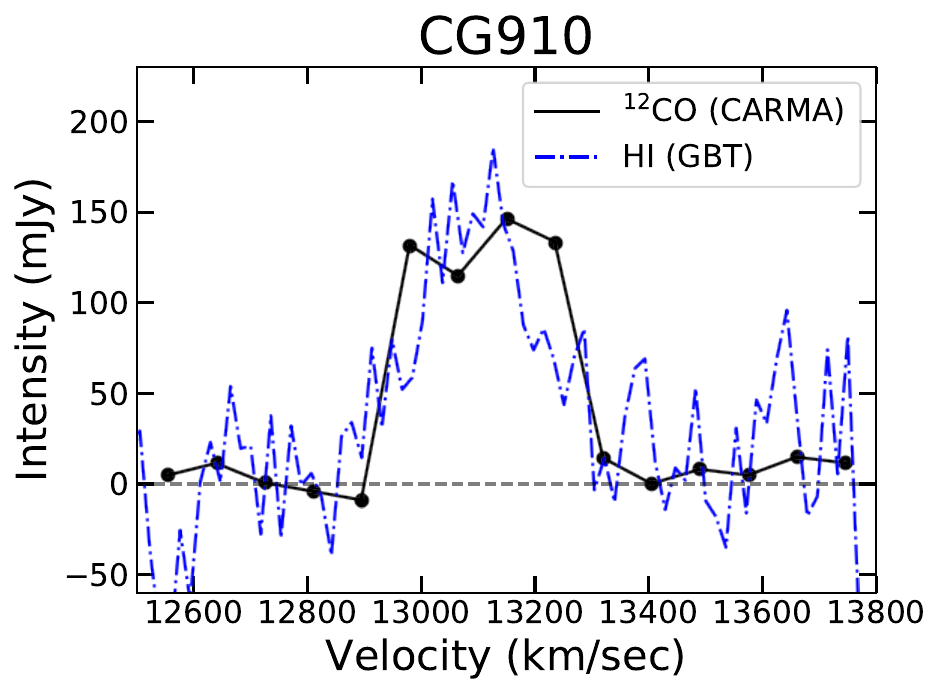}
\caption{The \hi~ emission taken from GBT observations in blue. The CARMA spectrum in black is overplotted. The HI flux is scaled by a factor of 80 for comparison.}\label{fig:av_HI}
\end{figure}

\subsection{The \hi~ line emission and mass}
 \begin{table}
\centering
\caption{{Line-fluxes and masses of the molecular and atomic gas}}
\label{tab:flux_mass}
\renewcommand{\arraystretch}{1.1}
\begin{tabular}{cccc} \hline
  &  Line-flux  &  Mass   \\ 
  &    (Jy \kms)  & (10$^{9}$ M$\odot$) \\
\hline
H$_{2}$  (CARMA) &  45$\pm$4 & 12.0$\pm$1.1 \\ 
H$_{2}$  (IRAM) &  13.8$\pm$3.1 & 3.7$\pm$0.8$\dagger$ \\
\hi~ (GBT) & 0.4$\pm$0.1  & 3.1$\pm$0.8 \\
\hline
\end{tabular}
\renewcommand{\arraystretch}{1.1}
\\
$^{\dagger}$ derived from \cite{sage.etal.1997} \\
\end{table}
The spectral profile for \hi~ emission obtained using GBT observations is averaged over the three days of observations and is shown in Fig. \ref{fig:av_HI} in blue. The emission peaks at the systematic velocity of the galaxy, which is 13134 \kms. We also over-plotted the CARMA CO profile in black to compare the molecular and atomic gas emission from the galaxy. The main component of the \hi~ spectral profile is in good correlation with CO emission at the velocity of $\sim$ 13134 \kms, suggesting that the cold gas in CG 910, i.e. both \hi~ and H$_2$, are closely associated and rotate together in the galaxy disk. The peak \hi~ flux for the main component is 2.2 mJy, and the exact systematic velocity obtained from the Gaussian fitting is 13118 \kms. The line width of \hi~ derived from the Gaussian fitting is $\sim$ 200 \kms. 
The \hi~ mass was estimated using the relation \citep{Roberts1962}:
\begin{equation}
\centering
    M(HI) = 1.4\times 2.36\times 10^{5}\times D_{Mpc}^{2}\times S_{\nu}  M_{\odot}
\end{equation}
where $S_{\nu}$ is the integrated flux density in units of Jy \kms and D$_{Mpc}$ is the luminosity distance as 188 Mpc. A factor of 1.4 has been included to correct for the presence of Helium. To obtain the integrated flux density, we determined the standard deviation of the flux in the emission-free channels and subtracted it from the main \hi~ spectrum. Table \ref{tab:flux_mass} shows the integrated flux-density of \hi~, which is 0.4$\pm$0.1 Jy \kms. The mass error is also determined using the standard deviation of the emission-free channels. Hence, the derived mass of the main component of the galaxy is estimated as $3.1\pm0.8\times10^{9}$ M$_{\odot}$. \cite{2012AJ....144...16K} studied a sample of 41 void galaxies with their \hi~ masses in the range of $1.7\times10^{8}$ to $5.5\times10^{9}$ M$_{\odot}$. Our estimation is consistent with their sample. However, the atomic mass of CG 910, M(\hi~), is smaller than the molecular mass M(H$_{2}$), which will be discussed further in a later section.

\subsection{Star formation efficiency and  Specific star formation rate}
The star formation efficiency is calculated as SFE = SFR/M(H$_{2}$), where the star formation rate (SFR) is calculated using the H$_{\alpha}$ flux from the MANGA map. Using SFR of 0.33 M$_{\odot}$ yr$^{-1}$ and M(H$_{2}$) of 12$\times10^{9}$ M$\odot$, we calculated SFE=$0.27\times 10^{-10}$ yr$^{-1}$. We also calculated the specific star formation rate, sSFR, as SFR/M$_{\star}$, where M$_{\star}$ is the stellar mass of the galaxy.  Using the archival stellar mass of $\sim21.5\times$10$^{9}$ M$\odot$, we calculated the sSFR=$0.07\times10^{-10}$ yr$^{-1}$ (log sSFR=-10.82). 

\section{Discussion}

We have studied the cold gas distribution (H$_2$ and \hi~) in the Bo\"otes void galaxy CG 910, which had previously been detected in CO(1--0) line emission using the single-dish IRAM telescope \citep{sage.etal.1997}. The results suggest that the molecular gas distribution is lopsided and asymmetrical around the nucleus. The gradient in the velocity dispersion map is also similar to the direction of the extended gas morphology in the CO intensity map, suggesting that the gas distribution is disturbed. Molecular gas or CO maps of void galaxies are rare, and this is possibly the first CO map of a void galaxy. The molecular gas mass in the inner disk is also surprisingly high (M(H$_2$)=$12\times10^{9}M_{\odot}$) compared to normal galaxies \citep{obreschkow.rawlings.2009}. Also, void galaxies are, in general, \hi~ dominated, but in this case, we find that the \hi~ gas mass is smaller than the H$_{2}$ mass. This is surprising but not completely unusual \citep{beygu.etal.2013}. Table \ref{tab:flux_mass} shows the details of the detected fluxes and gas masses in CG 910.  

In normal environments such as galaxy clusters, a lower M(HI) can be due to gas stripping associated with frequent galaxy interactions and mergers as the outer lying HI gas is more easily stripped compared to the H$_2$ gas that lies deep inside the disk. So, the main reason for M(HI)<M(H$_2$) must be a slow SFR compared to the rate of conversion of HI to M(H$_2$). To understand this further, we estimate the H$_2$ gas depletion time, t$_{dep}$ (H$_{2}$)= M(H$_2$)/SFR, which is the time scale required for a galaxy to consume the whole molecular gas mass at the current star formation rate (SFR). With an M(H$_2$) of $12\times10^{9}$ M$_{\odot}$ and SFR of 0.33 M$_{\odot}$ yr$^{-1}$, we find that t$_{dep}$(H$_{2}$)$\sim$36 Gyr.  The atomic gas depletion time scale, t$_{dep}$(HI) = M(HI)/SFR, is the time scale over which the atomic mass is converted to stars at a given star-formation rate, and we estimate the t$_{dep}$(HI) $\sim$9.27 Gyr. This time scale is high compared to star-forming galaxies, for example, in the xCOLD GASS survey \citep{2017ApJS..233...22S}, but is approximately close to the quenched galaxies that have t$_{dep}$ (HI)>10 Gyr \citep{2021ApJ...918...53G}. This value suggests that with its moderate SFR, CG 910 is forming stars over a relatively long time scale, thereby supporting the scenario of slow star formation in voids. Recent studies on void galaxies also suggest that the gas in voids is slowly assembled \citep{2022A&A...658A.124D,2023Natur.619..269D}.

Recent studies of massive star-forming galaxies (M$_{\star}$ =10$^{9}$-10$^{11.5}$ M$_{\odot}$) in the xGASS survey \citep{2018MNRAS.476..875C}, and the ALMaQUEST \citep{2024SCPMA..6799811Y} survey suggest that the variations in the molecular-to-atomic gas fraction are mostly driven by changes in the HI reservoirs, which are dependent on stellar mass surface densities. Although CG 910 is also a massive galaxy, the stellar and gas masses (M(H$_{2}$) or M(HI)) are almost comparable, whereas, for most star-forming disk galaxies, the gas mass is $<$10\% of the stellar mass. This suggests that the stellar mass surface density in CG 910 may not be large enough to provide the self-gravity required to support global disk instabilities such as bars and spiral arms. The latter dynamical processes are the main drivers of star formation in galaxies. A similar reasoning applies to low surface brightness galaxies, which have very diffuse stellar disks \citep{das.2013}. 
Thus, this study suggests that CG 910, like most void galaxies \citep{2011AJ....141....4K}, is slowly evolving and is still using up its gas content in star formation. 

\begin{table}
\centering
\caption{Photometric and derived star-formation properties of CG 910}
\label{tab:SFE}
\renewcommand{\arraystretch}{1.1}
\begin{tabular}{cccrr} \hline
\hline
Color (g-r) & 0.89 &  \\ 
Absolute magnitude (M$_r$) & -20.4  \\
M$_{\star}$ [M$_{\odot}$] & 21.5 $\times$ 10$^{9}$ (10.33) &   \\ 
 SFE [ yr$^{-1}$] & 0.27 $\times$10$^{-10}$(-10.56)   \\
 sSFR [yr$^{-1}$]  & 0.07 $\times$10$^{-10}$ (-10.82)  \\
 M$_{H_2}$/M$_{\star}$ & 0.56 (-0.25)  \\
 M$_{HI}$/M$_{\star}$  & 0.14 (-0.84)  \\
 M$_{H_2}$/M$_{HI}$ & 3.9 (0.58)  \\

\hline
\end{tabular}
\renewcommand{\arraystretch}{1.1}
\\
Values in parentheses are in the logarithmic scale. \\
\end{table}

The largest void galaxy survey so far, called CO-CAVITY, studied the molecular line emission for 200 galaxies using IRAM $^{12}$CO (1-0) and $^{12}$CO(2-1) observations \citep{2024arXiv241018078R}, and included the pilot study of 20 void galaxies \citep{2022A&A...658A.124D}. We compared the photometric and star formation properties of CG 910 ((M$H_{2}$, M$_{HI}$, SFE, sSFR, M$_{H_2}$/M$_{HI}$, M$_{HI}$/M$_{\star}$) with that of the sample. We find that CG 910 is redder (g-r =0.89) and more luminous (M$_{r}$ =-20.4) than most CO-CAVITY galaxies and lies at the upper brightness limit. \cite{2024arXiv241018078R} suggested that the star-formation properties of their void galaxy sample do not differ much from normal star-forming galaxies except at the high stellar mass end. The molecular hydrogen mass (log$M_{H_{2}}$) for the CO-CAVITY sample varies from 7.64 to 9.84. The authors derived the mean SFE in five different stellar mass bins (9.0 < log M$_{\star}$ < 11.5) and found that the SFE is constant with stellar mass and varies marginally from -8.94 to -8.99. The stellar and molecular mass of CG 910 lies at the higher end of the sample (log M$_{\star}$=10.33 and log M$_{H_{2}}$=10.079) when compared with the CO-CAVITY sample. The SFE for CG 91O (log SFE (yr$^{-1}$)=-10.56) is much lower than the overall distribution of the CO-CAVITY sample and falls in the higher stellar mass bin (M$_{\star}$ > 10$^{10.5}$ M$_{\odot}$). The mean sSFR ranges from -9.91 to -10.53, and so the sSFR of CG 910 (log sSFR=-10.82) is lower than the CO-CAVITY sample. For the molecular-to-atomic gas mass ratio, CG 910 (log M$_{H_2}$/M$_{HI}$= 0.58) is consistent with galaxies in the higher stellar mass bin (M$_{\star}$ > 10$^{10}$ M$_{\odot}$) as stated in \cite{2022A&A...658A.124D}. The atomic gas mass fraction (log M$_{HI}/M_{\star}$) of CG 910 is -0.84, lower than the average value in voids or for normal star-forming galaxies, but is consistent with the sample from CO-CAVITY.

One of the main reasons we studied this galaxy was to see if we could detect any ongoing gas accretion. This could be due to either interaction with close companion galaxies or cold gas accretion from the intergalactic medium via cosmic web filaments. A good way to search for gas accretion is to look for gas with abnormal velocities in the position velocity (PV) plot (Figure \ref{fig:pv_co}). Signatures of interaction may also be detected in molecular gas and appear as abnormal velocities in the PV plot. A good example of such interaction is the void galaxy Mrk 1477 or VGS31 \citep{2012AJ....144...16K}, which is part of a system of three interacting galaxies connected by a gas filament and shows significant CO emission \citep{beygu.etal.2013}. However, we did not detect any abnormal gas velocities in the CO observations. The PV plot shows only a marginal detection towards the southwestern side of the galaxy at a flux level of 0.05 Jy/beam. There is a hint of possible extended emission in the south-western direction of the galaxy, but considering the beam size of the CARMA data, this may or may not be real. Furthermore, optical images show no nearby companion galaxies that could trigger gas accretion, nor are there any signatures of a recent minor merger event in the galaxy. Thus, CG 910 shows no clear signatures of gas accretion in the CO data. 

We also searched for signatures of gas accretion in the obtained HI spectrum. It is now fairly well established from simulations that hot ionized gas from the circum-galactic medium (CGM) or the inter-galactic medium (IGM) in the cosmic web filaments cools and accretes onto galaxies \citep{nelson.etal.2016}. In observations, such cold gas accretion is expected to appear as filaments of cool gas near galaxies or low column density \hi~ in the halos of galaxies \citep{2008A&ARv..15..189S}. Such \hi~ filaments have been detected \citep{deblok.etal.2014} as well as low column density \hi~ gas \citep{das_sanskriti.etal.2020}. There may be some offset \hi~ in Figure~6 on the high-velocity side of the spectrum that represents gas accretion, but S/N is not good enough to draw any firm conclusions. 

\section{Conclusions}\label{sec:con}
We present CO and \hi~ observations of the galaxy CG 910, which lies in the  Bo\"otes void. The CARMA CO(1--0) observations reveal that the molecular gas has a mass of  M(H$_2$)=$12\pm1.1\times10^{9}M_{\odot}$ and is distributed in a disk of diameter 7 kpc. The CO velocity field shows a regularly rotating disk with a flat rotation velocity of 256 kms$^{-1}$. The CO velocity dispersion and the CO intensity peak in the centre suggest that the gas is concentrated in the bulge region. The single dish GBT observations of the \hi~ spectrum show that the \hi~ gas has a mass of M(\hi~)=3.1$\pm0.8\times10^{9}M_{\odot}$, is also centrally concentrated and has a distribution similar to the molecular gas. We have spatially resolved the molecular gas properties of a Void galaxy for the first time and attributed the low atomic gas mass fraction to the larger gas depletion timescale, confirming the slow evolution of the gas in voids. We do not find any substantial signatures of offset CO emission. There are hints of offset \hi~ emission in the GBT \hi~ spectrum, but the S/N is also not high. Hence, we do not detect any clear signatures of gas accretion in the void galaxy CG 910.

\medskip

\begin{acknowledgements}
The authors gratefully acknowledge the help of Prof. Stuart Vogel on the CARMA data and its analysis. 
This work is supported by the National Natural Science Foundation of China (NSFC) (Grant No. 11988101) and by the Alliance of International Science Organizations, Grant No. ANSO-VF-2021-01. The first author, E.S., is grateful to IIA for hosting her two-month visit and for the financial support from the Science and Engineering Research Board (SERB) MATRICS grant MTR/2020/000266 for that period. MD gratefully acknowledges the support of the Science and Engineering Research Board (SERB) Core Research Grant CRG/2022/004531, and the Department of Science and Technology (DST) grant DST/WIDUSHI-A/PM/2023/25(G) for this research. This material is based upon work supported by the Green Bank Observatory, which is a major facility funded by the National Science Foundation and operated by Associated Universities, Inc. The paper is also based on observations done with the Combined Array for Research in Millimeter Astronomy (CARMA) which was funded by the National Science Foundation.

The study has made use of the NASA Extragalactic Database (NED) and SDSS survey. Funding for the Sloan Digital Sky  Survey IV has been provided by the  Alfred P. Sloan Foundation, the U.S.  Department of Energy Office of  Science, and the Participating  Institutions. SDSS-IV acknowledges support and 
resources from the Center for High-Performance Computing  at the  University of Utah. The SDSS  website is \url{www.sdss4.org}. SDSS-IV is managed by the 
Astrophysical Research Consortium  for the Participating Institutions  of the SDSS Collaboration including  the Brazilian Participation Group,  the Carnegie Institution for Science, Carnegie Mellon University, Center for Astrophysics | Harvard \& Smithsonian, the Chilean Participation  Group, the French Participation Group,  Instituto de Astrof\'isica de  Canarias, The Johns Hopkins  University, Kavli Institute for the  Physics and Mathematics of the 
Universe (IPMU) / University of  Tokyo, the Korean Participation Group, 
Lawrence Berkeley National Laboratory,  Leibniz Institut f\"ur Astrophysik 
Potsdam (AIP),  Max-Planck-Institut  f\"ur Astronomie (MPIA Heidelberg), 
Max-Planck-Institut f\"ur  Astrophysik (MPA Garching),  Max-Planck-Institut f\"ur Extraterrestrische Physik (MPE), National Astronomical Observatories of 
China, New Mexico State University,  New York University, University of Notre Dame, Observat\'ario Nacional / MCTI, The Ohio State  University, Pennsylvania State  University, Shanghai Astronomical Observatory, United Kingdom Participation Group, Universidad Nacional Aut\'onoma de M\'exico, University of Arizona,  University of Colorado Boulder,  University of Oxford, University of Portsmouth, University of Utah,  University of Virginia, University 
of Washington, University of  Wisconsin, Vanderbilt University, and Yale University. This study has made use of Astropy, a community-developed core Python (\url{http://www.python.org}) package for Astronomy (Astropy Collaboration et al. 2013).

\end{acknowledgements}

\bibliographystyle{aa}
\bibliography{ref}

\end{document}